\documentclass[aps,superscriptaddress,showpacs,nofootinbib,eqsecnum]{revtex4-1}

\usepackage{xcolor}
\usepackage{epsfig} 
\input{epsf}
\usepackage{natbib} 
\usepackage[colorlinks=true,citecolor=blue,linkcolor=blue,urlcolor=blue]{
hyperref}
\setcitestyle{square, numbers, comma, sort&compress}

\def\be{\begin{equation}}
\def\ee{\end{equation}}
\def\bea{\begin{eqnarray}}
\def\eea{\end{eqnarray}}

\newcommand{\lattice}{Aoki2006, *Aoki_2009, *Borsanyi2010, *PhysRevD.85.054503}
\newcommand{\sign}{deForcrand:2010ys,*Aarts_2016}
\newcommand{\njl}{njl,*njl2,*BUBALLA2005205}
\newcommand{\QMM}{QMM1,*PhysRevD.97.034022}
\newcommand{\HTLpt}{PhysRevLett.83.2139, *PhysRevD.61.074016}
\newcommand{\HTLptL}{PhysRevLett.104.122003, *Andersen2010, *Andersen2011}
\newcommand{\HTLptDense}{Mogliacci2013, *PhysRevD.89.061701, *Haque2014}
\newcommand{\pms}{PhysRevD.23.2916,*STEVENSON1982472}
\newcommand{\spt}{KARSCH199769,*PhysRevD.63.105008,*PhysRevD.64.105012}
\newcommand{\lde}{Yukalov1976, *CASWELL1979153, *HALLIDAY1979421, 
*PhysRevA.34.5080, *JONES1990492, *NEVEU1991242, *doi:10.1063/1.529320, 
*PhysRevD.45.1248,*Yamada1993,*SISSAKIAN1994381,*PhysRevD.57.2264, 
*KLEINERT199874}
\newcommand{\coefGN}{doi:10.1142/S0217751X94000285,*GRACEY1992293}
\newcommand{\Kastening}{PhysRevD.54.3965,*PhysRevD.57.3567}
\newcommand{\coefQCD}{VERMASEREN1997327,*CZAKON2005485,*CHETYRKIN2005499}

\begin{document}

\title{Renormalization group improved pressure for cold and dense QCD}

\author{Jean-Lo\"{\i}c Kneur}\email{jean-loic.kneur@umontpellier.fr}
\affiliation{Laboratoire Charles Coulomb (L2C), UMR 5221 CNRS-Universit\'{e} de Montpellier, 34095 Montpellier, France}  
\author{Marcus Benghi Pinto} \email{marcus.benghi@ufsc.br}
\affiliation{Departamento de F\'{\i}sica, Universidade Federal de Santa
  Catarina, Florian\'{o}polis, SC 88040-900, Brazil}
\affiliation{Department of Physics, University of Colorado, Boulder, CO 80309, USA}
\author{Tulio Eduardo Restrepo} \email{tulio.restrepo@posgrad.ufsc.br}
\affiliation{Departamento de F\'{\i}sica, Universidade Federal de Santa
  Catarina, Florian\'{o}polis, SC 88040-900, Brazil}
 
 \begin{abstract} 
 We apply the renormalization group optimized perturbation theory (RGOPT) 
 to evaluate the QCD (matter) pressure  at the  two-loop level considering three flavors of massless quarks 
 in a dense and cold medium.  Already at leading order ($\alpha_s^0$), which builds on the 
 simple one loop (RG resummed) term,
 our technique provides a non-trivial non-perturbative approximation which is completely renormalization group invariant. 
 At the next-to-leading order  the comparison between the  RGOPT and the perturbative QCD predictions shows that the former method 
 provides results which are in better agreement with the state-of-the-art  {\em higher order} perturbative results, 
 which include a contribution of order $\alpha_s^3 \ln^2 \alpha_s$.  At the same time one also observes that the RGOPT 
 predictions are less sensitive to variations of the arbitrary $\overline {\rm MS}$ renormalization scale  
 than those obtained with perturbative QCD.  These results indicate that the RGOPT provides an  efficient resummation scheme 
 which may be considered as an alternative to lattice simulations at high baryonic densities.
\end{abstract}

 
\maketitle
\section{Introduction}

First principles evaluations aiming to describe the properties of strongly interacting matter at finite temperatures 
and/or baryonic densities are highly complicated by the 
inherent non-linear and non-perturbative characteristics displayed by quantum chromodynamics (QCD). 
Nevertheless, at least  in regimes of vanishing baryonic densities which concerns high energy heavy ion collisions, 
this fundamental theory can nowadays be successfully described by numerical 
lattice simulations (LQCD) \cite{\lattice}. 
However, the famous sign problem \cite {\sign} for nonzero chemical potential is 
still preventing the method to be 
reliably applied to regimes of intermediate temperatures and baryonic densities which are relevant to
experiments such as the {\it Beam Energy Scam} (BES) at the Relativistic Heavy Ion Collider (RHIC) as well as 
CBM at FAIR and NICA at JINR which aim to locate the eventual QCD critical end point. At the same time the knowledge
of an equation of state (EoS) that faithfully describes the cold and dense regime is necessary for an accurate 
description of  compact stellar objects. Unfortunately, for the reasons alluded above, LQCD cannot yet furnish such 
an EoS so that in general the problem is partially circumvented in different ways such as by using chiral effective 
theories (CET) \cite {CET}   at low densities and perturbative QCD (pQCD) \cite {pQCD} at  ultrahigh densities. 
As an alternative to these two (first principles) analytical approaches one may use effective quark models such as 
the MIT bag model \cite {mit}, the Nambu--Jona-Lasinio model (NJL) \cite {\njl}, 
as well as the quark-meson model 
(QMM) \cite {\QMM} among others. In principle, pQCD applications should be 
carried out at  extremely high densities 
where the asymptotic freedom property assures that the QCD coupling, $\alpha_s$, is small enough to justify the use
of such an approximation. A seminal pQCD work by Freedman and McLerran \cite {mclerran1, mclerran2} has provided the
next-to-next-to-leading order (NNLO) pressure for massless quarks at vanishing temperatures and finite chemical 
potentials. The result has then been further refined so as to  include thermal effects \cite {thermal1,thermal2} and
finite quark masses \cite{edpaul, paul3L,mass} apart from being rederived in a way compatible  with the more modern 
${\rm {\overline {MS}}}$ renormalization scheme \cite {LambdaMSbar}.  After more than four decades, a new perturbative 
order has recently been evaluated in Ref. \cite {paulPRL}, where the authors have determined the coefficient of the 
leading-logarithmic contribution at N3LO: $\alpha_s^3 \ln^2 \alpha_s$. Since the leading-logarithm soft contribution 
at N3LO evaluated in that work  gives a negligible correction to the NNLO it was concluded that   using pQCD result 
as an {\it ab-initio} input in calculations of the properties of neutron stars \cite{paul3L,props1,props2,props3,props4}  as
well  as  simulation  gravitational-wave  signals from neutron-star mergers is well justified. Nevertheless, for our
present purposes it is also important to remark that in the evaluations performed in Ref. \cite {paulPRL} the authors 
have chosen the ${\rm {\overline {MS}}}$ arbitrary scale ($M$) to be $2\mu$ (where $\mu$ is the quark chemical 
potential). However, it is well known that physical observables evaluated with standard  perturbation theory,  
as well as those obtained with resummation methods such as hard thermal loop perturbation 
theory (HTLpt) 
\cite{\HTLpt,PhysRevLett.104.122003,*Andersen2010,*Andersen2011,Mogliacci2013,
*PhysRevD.89.061701,*Haque2014}, can be very sensitive to 
renormalization scale variations.
Moreover, it has been observed notably at finite 
temperature \cite{pQCD,blaizot,Mogliacci2013,*PhysRevD.89.061701,*Haque2014} 
that the latter 
scale sensitivity even {\it increases} when successive terms in the weak-coupling expansion are considered, 
which is an odd result as far as thermodynamical observables, such as the pressure, are concerned. 
At vanishing temperatures and finite baryonic densities, the scale dependence of the QCD pressure 
at NLO and NNLO has been 
explicitly investigated in Ref. \cite {paul3L}. The results show that the pressure has a rather large 
renormalization scale dependence, especially below the quark chemical potential $\mu \sim 1 \, {\rm GeV}$, which 
corresponds to a baryon density $\sim 10^2\, \rho_0$ where $\rho_0 \sim 0.16\, {\rm fm}^{-3}$ represents 
the nuclear mass density. Such dependence indicates that the eventual non-perturbative effects remain quite important 
in the lower density range relevant to neutron stars. We remark that the renormalization scale dependence is even 
worse with the HTLpt resummation at finite $T$, where the calculations have been pushed to the three loop level, 
predicting results which agree with LQCD but only when the central scale $M=2\pi T$ is used at 
$\mu=0$ \cite {\HTLptL}, while 
exhibiting a very large variation from $M=\pi T$ to $M= 4\pi T$. 
This is a clear indication that renormalization group (RG) properties have not been properly addressed 
in the perturbative and HTLpt resummed evaluations. The results displayed in 
Ref. \cite {\HTLptDense} show 
that this unfortunate situation persists at finite densities and finite temperatures  when the scale is varied 
around $M=2\pi \sqrt{ T^ 2 +\mu^2/\pi}$.\\

In the present work we consider an alternative resummation method which incorporates RG properties to evaluate
the QCD pressure at $T=0$ and finite $\mu$ values at the two loop level. This technique, which provides
non-perturbative approximations, has been dubbed RGOPT 
(renormalization group optimized perturbation theory), and can be viewed as an extension of the standard optimized 
perturbation theory (OPT) \cite {\pms, opt} and the screened perturbation theory 
(SPT)  \cite {\spt}  (both related 
to the  so called linear $\delta$ expansion (LDE) \cite {\lde}). Remark also 
that the HTLpt can be seen as  the 
gauge-invariance compatible version of the OPT/SPT. Initially the RGOPT was
employed \cite{JLGN} at vanishing temperatures and densities
in the Gross-Neveu (GN) model \cite {gn}. Then, it has been applied to QCD at $T=\mu=0$ to evaluate the basic scale 
$\Lambda_{\overline{\rm MS}}$\cite{JLLambda,JLalphas} (equivalently the coupling $\alpha_s$), 
predicting values compatible with the world average\cite{PDG2018}. The method has also been used to 
derive an accurate value of the quark condensate \cite{JLcond}. 
More recently, it has been applied to the scalar $\lambda \phi^4$ theory \cite{prdphi4,prlphi4}, 
as well as to the non-linear sigma model (NLSM) \cite {nlsm}, producing results that show its 
compatibility with control parameters such as the temperature. 
 The present paper is the first RGOPT application to (cold) in-medium QCD, for  
a non-zero  chemical potential, so that one can analyze how the method performs in the regime of finite baryonic 
densities. The latter, despite being currently largely unaccessible to LQCD, 
is of utmost importance to the description of compact stellar objects. 
Our goal is twofold: first, we would like to check how our approach compares
with the N3LO pQCD results recently obtained in Ref. \cite{paulPRL}. Second, we aim to show how the 
scale dependence within the predicted pQCD pressure can be significantly reduced when the evaluations 
are performed within the RGOPT, a generic feature of the method.
The paper is organized as follows. As a warm up, in the next section we  review the RGOPT approach 
illustrating it with the $d=1+1$ massless Gross-Neveu model at $T=\mu=0$.  In Sec. III the method is 
used to evaluate the quark contribution to the QCD pressure at vanishing temperatures and finite densities up to 
the (RG optimized) NLO two-loop level. The optimization procedure and numerical results are presented  in Sec. IV. 
Then in Sec. V we present our conclusions and perspectives.

\section{Reviewing the RGOPT with the GN model}\label{secGN}
The  RGOPT belongs to a class of variational methods, reminiscent of the traditional Hartree approximation, 
which are particularly suitable to tackle  infrared problems that plague massless theories.
In this section  the main steps of the approach will be recalled by 
performing a simple lowest order application to the massless Gross-Neveu model (GN) in two dimensions. 
More details and applications of the method can be found in Refs.\cite{JLGN,JLalphas,JLcond, prdphi4, prlphi4, nlsm}. 
The GN model is described by the Lagrangian density for a fermion field with $N$ components given by \cite{gn}
\begin{equation} 
{\cal L}_{GN} = 
\overline{\psi} \left( i \not\!\partial\right) \psi  + \frac {g_{GN}^2}{2} ({\overline {\psi}} \psi)^2\;.
\label{GN} 
\end{equation} 
The theory described by Eq. (\ref {GN}) 
is invariant under the  transformation $\psi \to \gamma_5 \psi$ displaying a discrete chiral symmetry (CS) in 
addition to having a global  $O(2N)$ flavor symmetry. This simple renormalizable model has important common 
features with QCD, such as asymptotic freedom and a dynamically generated mass gap, among others. It is exactly 
solvable in the large-$N$ limit, and for arbitrary $N$ values the exact mass gap (at vanishing temperature) has been  
obtained \cite{GNTBA} from Bethe {\it ansatz} methods. 
This allows to confront other non-perturbative approximation schemes that can include finite $N$ corrections 
(such as the RGOPT) to either the large- or finite-$N$ known results. \\
For convenience let us first rescale the four-fermion interaction as $ g_{GN}^2  = g {\pi}/N$.  
To implement the RGOPT requires first to deform the interaction terms of Eq.~(\ref{GN}) 
by introducing a Gaussian interpolating (mass) term and rescaling the coupling:
in the case of a {\it massless} theory the RGOPT prescription is
\begin{equation}
{\cal L}_{GN}^{RGOPT}= {\cal L}_{GN}(g  \to \delta g)  - m(1-\delta)^a \;,
 \label{LintGN}
\end{equation}

where $\delta$ is a book-keeping parameter interpolating between the free massive ($\delta=0$) and 
interacting massless ($\delta=1$) theory.
Remark that setting  $a=1$ in Eq.(\ref{LintGN}) gives  
simply the ``added and subtracted'' variational mass 
prescription as adopted in the standard OPT/SPT/LDE 
\cite{opt,KARSCH199769, *PhysRevD.63.105008, *PhysRevD.64.105012, Yukalov1976, 
*CASWELL1979153, *HALLIDAY1979421, *PhysRevA.34.5080,*JONES1990492, 
*NEVEU1991242, *doi:10.1063/1.529320, *PhysRevD.45.1248, *Yamada1993, 
*SISSAKIAN1994381, *PhysRevD.57.2264, *KLEINERT199874}. In contrast a crucial 
feature of the RGOPT is
 to {\em determine} \cite{JLGN,JLLambda,JLalphas} the exponent $a$ from renormalization group consistency, 
 giving generally $a\ne 1$, as we will recap below.
Note that for the original massless model, the (free) propagator would normally be 
$S_F(p) = i ( \not\! p )^{-1}$,  while within the OPT or RGOPT approaches, any perturbative evaluations
are first performed with a nonvanishing mass, thus providing an infrared regulator mass $m$, prior to 
the substitution Eq.~(\ref{LintGN}) (the latter being most conveniently performed after a standard perturbative
renormalization).\\ 
In the sequel of this section, to present a clearer overall picture of the approach we also restrict ourselves 
to the $T=\mu=0$ case,
since the main RGOPT features that we aim to recap are essentially determined by RG properties (thus
by the renormalization aspects of the $T=\mu=0$ part only). Once such RG properties are fixed, including 
the thermal and/or chemical potential contributions at a given order  
amounts to perform consistently the very same modifications as implied by Eq.~(\ref{LintGN})
within those perturbatively calculated (massive) contributions.\\ 
We then start by evaluating the leading order ${\cal O}(g^0)$ perturbative 
vacuum energy  of the massive GN-model (more generally we could consider the 
pressure, with $T\ne 0$ and/or $\mu \ne 0$)  
\begin{equation}
\frac{{\cal E}^{PT}_0}{N} = i\int\frac{d^2 p}{\left(2\pi\right)^2} \ln\left(p^2-m^2\right)+ {\cal O}(g ) \;.
\end{equation}
After renormalizing in the ${\rm {\overline {MS}}}$-scheme (which at this lowest order amounts to simply 
a vacuum energy counterterm), one obtains 
\begin{equation}
 \frac{{\cal E}^{PT}}{N}=-\frac{m^2}{2\pi}\left(\frac{1}{2}-L_m\right)+ \mathcal{O} \left(g \right)\;,
 \label{Ppt}
 \end{equation}
 where $L_m = \ln (m/M)$ and $M$ is the arbitrary $\overline {\rm MS}$ renormalization scale. 
Next consider the RG operator
\begin{equation}
M \frac{d}{d M}= M \frac{\partial}{\partial M} + \beta(g )\frac{\partial}{\partial g } - 
\gamma_m(g ) m \frac{\partial}{\partial m} \;,
\label{RG}
\end{equation}
with the normalization conventions for the RG coefficients~\cite {\coefGN}:
\begin{equation}
\beta(g )= -b_0 {g ^2} - b_1  {g ^3}+ {\cal O}(g ^4)\;,
\end{equation}
\begin{equation}
\gamma_m(g )=\gamma_0  {g } + \gamma_1^{\overline {\rm MS}} {g ^2}+
{\cal O}(g ^3) \;,
\end{equation}
where $b_0= 1 - 1/N$, $b_1=-b_0/(2N)$,  $\gamma_0=1-1/(2N)$, and 
$\gamma_1^{\overline {\rm MS}}= -\gamma_0/(4N)$.  
The next step is to realize that Eq.~(\ref{Ppt}) is not perturbatively RG-invariant:
applying Eq.(\ref{RG}) to this expression gives a remnant term of {\em leading} order:
$ M d {\cal E}/d M = -m^2 N/(2 \pi) \ne O(g )$. 
However this rather well-known problem of a massive theory can be solved most
conveniently by simply subtracting a (zero point) finite term in order to restore a RG invariant  perturbative 
vacuum energy\footnote{Alternatively one finds the very same results
by requiring RG invariance at the level of bare expressions\cite{JLGN}.}, that lead to the RG 
invariant (RGI) observable \cite{JLalphas}
\begin{equation}
{\cal E}^{RGI} = {\cal E}^{PT} - \frac{m^2}{g } s_0 \;.
\label{s0sub}
\end{equation} 
Now requiring Eq. (\ref{s0sub}) to satisfy the RG equations  fixes the $s_0$ coefficient to
\begin{equation}
s_0 = - N\,[2{\pi}(b_0-2 \gamma_0)]^{-1} =\frac{N}{2\pi}\;.
\end{equation}
The procedure is easily generalized most conveniently when taking higher perturbative order contributions into account 
by considering a perturbative subtraction
\begin{equation}
 - m^2 \sum_{k\ge 0} s_k g^{k-1} \;,
\label{general}
\end{equation} 
where the successive $s_i$ coefficients are fixed by requiring perturbative RG invariance, 
consistently including higher orders within the RG Eq. (\ref{RG}). This perturbative 
RG invariance restoration is of course not specific to the $d=1+1$ GN model
but more generic for any massive model, thus also in particular in four dimensions 
(see e.g.  \cite{\Kastening} for high order analysis in the $\phi^4$ theory). 
Now, incorporating those necessary subtraction terms, in order to start from a perturbatively RG invariant 
quantity, is an important necessary step prior to the specific RGOPT modification implied by Eq.(\ref{LintGN}).
Next, performing the replacements Eq.~(\ref{LintGN}) within a perturbative expression like (\ref{s0sub}) 
and doing a power expansion to order-$\delta^k$,
one aims to recover formally the massless limit, $\delta\to 1$. But the latter (re)expansion 
leaves a remnant dependence on the (arbitrary) mass $m$ at any finite $\delta^k$ 
order, since the expression was initially perturbative. Indeed, 
applying the RGOPT replacements, at lowest $\delta^0$ order, to Eq. (\ref{s0sub}) 
gives
\begin{equation} 
 \frac{{\cal E}^{RGOPT}}{N}=-\frac{m^2}{2\pi}\left (\frac{1}{2}-L_m\right) - \frac{m^2}{N g }(1-2a)s_0 \;.
 \label{P0d0}
\end{equation}
Now a crucial step is to realize that the resulting modified perturbative expression, Eq. (\ref{P0d0}), 
spoils the RG invariance in general, 
in particular for the simplest ``added and subtracted mass" prescription $a=1$, 
due to the drastic modification of the mass dependence.
In contrast, the idea is to determine the interpolation exponent $a$ in Eq.(\ref{LintGN}) by requiring rather 
the {\it reduced } RG equation\cite {JLGN} to hold:
\begin{equation}
f_{RG} \equiv M \frac{\partial {\cal E}^{RGOPT} }{\partial M} + \beta(g)\frac{\partial 
{\cal E}^{RGOPT}}{\partial g } \equiv 0 \;,
\label{red}
\end{equation}
in consistency with the {\em massless} limit being seeked out.
This uniquely fixes the exponent as 
\begin{equation}
 a=\frac{\gamma_0}{b_0}\; ,
 \label{aGN}
\end{equation}
a generic result also for other theories \cite {JLalphas,JLcond,prdphi4,prlphi4,nlsm}.  
Moreover, the same value of $a$ is taken also 
when considering higher orders of the $\delta$-expansion, keeping the simple interpolating form of Eq.~(\ref{LintGN}),
since this exponent is universal (renormalization scheme independent).  

Thus substituting $a=\gamma_0/b_0$ into Eq. (\ref {P0d0}) leads to the RGOPT  lowest order result
\begin{equation} 
 \frac{{\cal E}^{RGOPT}}{N}=-\frac{m^2}{2\pi}\left (\frac{1}{2}-L_m\right) + \frac{m^2}{2 {\pi}  g  b_0} \;.
 \label{ErgoGN}
\end{equation}
It is important to note that already at this lowest order the RGOPT-modified subtraction term clearly 
brings dynamical (RG) information through $g$ and $b_0$ apart from finite $N$ contributions 
(since $b_0=1-1/N$) to an otherwise trivial (free) vacuum energy. 
At this lowest order the final step consists in fixing the parameter $m$, still arbitrary at this stage,
with an optimization prescription (MOP), similar to the so-called principle of minimal sensitivity 
(PMS) \cite {\pms}, defined by the stationarity condition 
\begin{equation}
f_{MOP} = \frac{\partial {\cal E}^{RGOPT}}{\partial m} \equiv 0 = \frac{m}{\pi} 
\left( \frac{1}{b_0\,g} +L_m \right) \,.
\label{pms}
\end{equation}

Apart from the trivial result ${\overline m}=0$, one obtains
\begin{equation}
{\overline m}= M \exp[ -{1}/(b_0\, g)] \; ,
\end{equation}
which is clearly non-perturbative and explicitly RG invariant. Substituting $\overline m$ within
Eq.(\ref{ErgoGN}), the vacuum energy is also RG invariant, and immediately gives the correct large-$N$
result, as was observed in ref.\cite{JLalphas}. 
To better appreciate these RGOPT features let us now compare this result with 
those obtained by the standard OPT/SPT as well as the large-$N$ approximations.  
As shown in Ref. \cite{prd}, at order-$\delta^0$  the standard OPT/SPT vacuum energy (or equivalently pressure) 
has no information about the interactions since it is $g$-independent. Therefore the first non trivial contribution 
arises at next order-$\delta$ (two loop level) and by applying the MOP criterion one fixes 
 the mass to
 \begin{equation}
{\overline m}_{OPT}= M \exp[ -{1}/(g  \gamma_0)] \; ,
\end{equation}
which {\it is not} RG invariant. As for the $1/N$ expansion, the first non trivial
contribution appears at order-$N^0$ (the large-$N$ limit, LN)  whose gap equation yields 
the well known non-perturbative mass gap
\begin{equation}
m^{LN}= M \exp(-{1}/g ) \; .
\end{equation}
At this point a remarkable property of the RGOPT procedure over LN and standard OPT should be clear: 
it does produce a scale invariant non-perturbative result, which incorporates finite $N$ corrections, 
already at the one loop level. The same properties
hold whenever adding in-medium contributions, because
the latter are not affecting those RG properties which essentially rely on the vacuum contributions. 
Moreover, for $N \to \infty$ the RGOPT also reproduces the ``exact" LN 
result, a consistency check of the reliability of the method. 
We point out that the latter property 
is also observed within the standard OPT since, as a particularity of the GN model, 
$\gamma_0 = b_0 \equiv 1$ at large-$N$~\footnote{The fact that, for the GN as well as other theories, 
the OPT type of method does reproduce the $N\to\infty$ limit was observed long ago \cite {npb}.}.  
The LN limit is also reproduced when considering in-medium effects: 
for example for the $\phi^4$ model, quite remarkably the lowest order RGOPT pressure reproduces 
correctly \cite{prdphi4} the (all-order) exact properties 
of the LN limit (that in the more standard large-$N$ derivation \cite{phi4N} involve the nontrivial 
resummation of ``daisy" and ``superdaisy" graphs, associated with plasmon infrared divergent contributions). 
Yet for more involved theories such as QCD, one does not expect the lowest 
$\delta^0$-order RGOPT to be a very realistic approximation in general. 
This is because it essentially relies on lowest order RG quantities,
while the other relevant contributions, e.g., in the pressure,
are essentially those from a free theory at this order. Accordingly it appears sensible to go at least 
to the NLO order to get numerically more realistic results \cite{JLalphas,JLcond,prlphi4}. 
As we will illustrate
in next sections  this will be the case also for the QCD in-medium thermodynamic quantities 
considered in this work.

We will not proceed further with the GN model but to conclude this section we mention 
that the RGOPT recipe generalization to higher orders is rather straightforward, as will be better
illustrated in the next sections with the in-medium QCD case. Once having implemented the relevant 
RG subtraction coefficients in Eq.(\ref{general}), one performs the RGOPT modification from Eq.(\ref{LintGN}) 
using the universal $a$ value, Eq. (\ref{aGN}), expanding this to $\delta^k$-order consistently with the 
original perturbative order considered, and taking the massless limit $\delta\to 1$.
Finally one uses the RG Eq.(\ref{red}) and (or) the MOP Eq.(\ref{pms})   
to obtain ``non-perturbative'' approximations, in the sense that the resulting RG-consistent {\em dressed}
mass is of order $\Lambda_{\overline{\text{MS}}}$ at $T=\mu=0$\cite{JLalphas}. At non-vanishing 
temperatures the dressed mass also acquires a thermal dependence, but keeping its RG properties
(see Refs. \cite {prdphi4,prlphi4,nlsm} for more detailed discussions). \\
Ideally one would aim to solve the two Eqs. (\ref{pms}), (\ref{red}) simultaneously to fix both a 
dressed running mass ($\overline m$) and a dressed running coupling ($\overline g $). 
However, as one proceeds to higher orders both equations often develop non linearities, 
so that an increasing number of solutions occur a priori, moreover not guaranteed to be all
real-valued. These unwelcome features are indeed common with the other related OPT/SPT
approaches.  But in the RGOPT, Eq.(\ref{aGN}) also crucially guarantees 
that the only acceptable solutions are those matching the standard perturbative behavior for $g\to 0$ at $T=0$, 
a simple criteria that most often selects a unique solution, even at the highest (four-loop) order
investigated so far \cite{JLalphas,JLcond}. 
Alternatively a less complete but often more handy RG compatible criterion requires to 
solving only the full RG Eq. (\ref{RG}), to fix the dressed mass 
${\overline m}(g )$. Next the coupling (not yet fixed at this stage) is naturally 
traded for the ordinary running coupling at the relevant perturbative order, instead of being more
non-perturbatively determined. Accordingly, the final physical quantities exhibit 
a more pronounced residual scale dependence, which can be interpreted as an estimate of the error 
introduced by this alternative procedure. 
Nevertheless, different applications have shown that this residual scale dependence is milder compared 
to the ones produced by standard PT and also by the related OPT/SPT approaches. 

\section{RGOPT evaluation of the QCD quark pressure }
Let us now apply the RGOPT to the three  flavor (dense matter) 
QCD up to order-$g$ (defining for convenience $g= 4 \pi \alpha_s$), in the limit of vanishing 
temperatures and finite baryonic densities with $\mu_s=\mu_u=\mu_d\equiv \mu$, 
which is the equilibrium condition for the massless case considered here. 
To thus treat properly the quark sector of QCD, the  RGOPT 
requires to deforming the theory by rescaling the coupling 
(consistently for every standard QCD interaction terms) 
and a  modified Gaussian interpolating (mass) term, following the prescription 
\begin{equation}
{\cal L}_{QCD}^{RGOPT}= {\cal L}_{QCD}\vert_{g \to \delta g} -m(1-\delta)^a {\overline \psi}_f \psi_f ,
 \label{Lint}
\end{equation}
where $f=u,d,s$ is flavor index.
The fermionic interpolating term proportional to $m$ is completely similar to the one previously discussed for
the GN model. Note carefully that in order to compare with 
Ref. \cite {paulPRL} in the present work we will investigate the case of vanishing {\em current} masses 
($m_u=m_d=m_s=0$), while $m$ in Eq. (\ref {PPTqcd}) above will become our variational mass upon implementing 
the RGOPT replacements, just as in the GN case illustrated in the previous section \ref{secGN}.
(Accordingly $m$ represents a generic mass identical for the three flavors,
in this initially $SU(3)$ flavor symmetric approximation.)
 As a parenthetical remark, in principle a more complete and rather similar treatment 
of the gluon sector is possible, by following the hard thermal loop (HTL) prescription
originally suggested by Braaten and Pisarski \cite {termHTLpt}, that essentially
introduces a gauge-invariant (non-local) effective Lagrangian properly describing a gluonic (thermal) 
``mass" term in the HTLpt approximation 
\cite{PhysRevLett.83.2139, *PhysRevD.61.074016, PhysRevLett.104.122003, 
*Andersen2010, *Andersen2011, Mogliacci2013, *PhysRevD.89.061701, *Haque2014}. 
 
However, in the present work, which deals only with the $T=0$ and $\mu \ne 0$ regime, we will apply the 
RGOPT to the quark sector only so that the gluon propagator, entering our evaluation at two-loop order, will 
be  the usual (massless) one used in purely perturbative QCD (thus also with standard QCD interactions 
with quarks once the appropriate $\delta\to 1$ limit is taken, after the $\delta$-expansion following
Eq.~(\ref{Lint}).  
This is justified by aiming to compare our results with the purely perturbative evaluation of the cold pressure
such as in Ref.\cite{paulPRL}, also since the HTL-modified Lagrangian is supposed 
to play a crucial role more essentially once considering high temperature effects. 
\begin{figure}[htb]
 \includegraphics{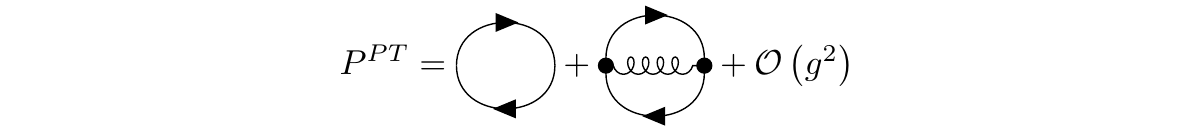}
\caption{Feynman diagrams contributing to the perturbative quark pressure up to order-$g$. }
\label{Fig1}
\end{figure}

To order-$g$ the relevant contributions are displayed in Fig. \ref {Fig1}. By combining the results 
of Ref. \cite {JLcond}  for the vacuum ($\mu=0$) contributions with those of Ref. \cite {edpaul} for the in-medium part one 
obtains the renormalized result 
\begin{eqnarray}
P^{PT}_{1,f}(\mu)&=&-N_c \frac{m^4}{8 \pi^2} \left(\frac{3}{4}-L_m\right)+\Theta(\mu^2-m^2)\,\frac{N_c}{12 \pi^2} 
\left [ \mu p_F\left (\mu^2 - \frac{5}{2} m^2 \right ) + \frac{3}{2} m^4 \ln ( \frac{\mu +p_F}{m} ) 
\right ]\nonumber \\
 &-& \frac{d_A \,g}{4\left(2\pi\right)^4} m^4 \left(3 L_m^2-4 L_m+\frac{9}{4}\right) 
 -\Theta(\mu^2-m^2)\, \frac{d_A\, g}{4\left(2\pi\right)^4} \left \{ 3 \left [m^2 \ln ( \frac{\mu +p_F}{m} ) 
 -\mu p_F\right ]^2 - 2 p_F^4 \right \}
 \nonumber \\
&-& \Theta(\mu^2-m^2)\, \frac{d_A\,g}{4\left(2\pi\right)^4} m^2  \left ( 4- 6 L_m \right ) 
 \left [ \mu p_F - m^2 \ln ( \frac{\mu +p_F}{m} ) \right ]   \;,
\label{PPTqcd} 
\end{eqnarray}
where $p_F=\sqrt{\mu^2 -m^2}$ is the Fermi momentum, $L_m= \ln(m/M)$, $d_A = N_c^2-1$, and $N_c=3$.  
Now, to turn the above pressure into a RG invariant quantity, as explained in previous section, 
we subtract a finite ``zero-point" contribution: 
\begin{equation}
P^{PT}_{1,f}(\mu) \to P^{RGI}_{1,f}(\mu)=P^{PT}_{1,f}(\mu) -m^4 \sum_k s_k g^{k-1}\; .
 \label{sub}
\end{equation}
 Since our evaluations are being carried up to two-loop,  
order-$g$, it suffices to determine the first two coefficients $s_0$ and $s_1$ from applying the RG  
 to the pressure (at $T=\mu=0$), with the appropriate QCD $\beta$ and $\gamma_m$ RG functions. 
 In our normalization conventions the QCD $\beta(g)$ and $\gamma_m(g)$ 
functions read
\begin{equation}
 \beta\left(g\equiv 4\pi\alpha_S \right)=-2b_0g^2-2b_1g^3+\mathcal{O}\left(g^4\right) \;,
 \label{betaQCD}
 \end{equation}
 and
 \begin{equation}
 \gamma_m\left(g\right)=\gamma_0g+\gamma_1g^2+\mathcal{O}\left(g^3\right)\;,
\end{equation}
where the coefficients  are \cite {\coefQCD}
\begin{eqnarray}
 b_0&=& \frac{1}{\left(4\pi\right)^2}\left(11-\frac{2}{3}N_f\right ),\\
 b_1&=& \frac{1}{\left(4\pi\right)^4}\left (102-\frac{38}{3}N_f\right ),\\
 \gamma_0&=&\frac{1}{2\pi^2}
 \end{eqnarray}
 and
\begin{equation}
 \gamma_1^{\overline {\rm MS}}= \frac{1}{8\left(2\pi\right)^4}\left (\frac{202}{3}-\frac{20}{9}N_f\right) \;.
\end{equation}
Applying the RG Eq.(\ref{RG}) to Eq. (\ref{PPTqcd}) and requiring the result
to vanish up to higher ${\cal O}(g^2)$ terms determine
the subtraction coefficients in Eq.~(\ref{sub}) to
\begin{equation}
 s_0=- N_c \left[(4\pi)^2 (b_0-2\gamma_0)\right]^{-1} \;,
 \end{equation}
 and
 \begin{equation}
 s_1=-\frac{N_c}{4}\left[\frac{b_1-2\gamma_1}{4(b_0-2\gamma_0)} -\frac{1}{12\pi^2}\right ]\,.
 \label{s1}
\end{equation}
Remark that the coefficients $s_k$, being determined solely 
from the vacuum contributions, do not depend on the mass nor on control parameters such 
as the temperature and chemical potential \cite {prdphi4, prlphi4, nlsm}. Next, to implement the 
actual RGOPT modification of interactions, we 
follow the substitution prescribed in Eq.(\ref{Lint}). 
Like in the GN case the next step is to fix the exponent $a$, by expanding to leading order-$\delta^0$ and 
requiring the resulting pressure to satisfy the {\it reduced} RG Eq.(\ref{red}), here applied to the QCD pressure.
As expected this can be checked to yield the universal exponent\footnote{Notice a trivial factor $1/2$
difference as compared to Eq.(\ref{aGN}) due to a convenient different normalization of $b_i$ in Eq.~(\ref{betaQCD}).}: 
\begin{equation}
 a= \frac{\gamma_0}{2 b_0} \;,
 \label{aQCD}
\end{equation}
 in agreement with 
previous RGOPT applications to QCD \cite{JLalphas,JLcond}. The  LO RGOPT pressure, per flavor, 
can then be written as 
\begin{equation}
  P^{RGOPT}_{0,f}(\mu)=
  N_c \frac{m^4}{\left(4\pi\right)^2 b_0 \, g} + 
  \frac{N_c}{12 \pi^2} \left [ \mu p_F\left (\mu^2 - \frac{5}{2} m^2 \right ) 
  + \frac{3}{2} m^4 \left ( L_\mu -\frac{3}{4} \right ) \right ]\;,
 \label{rgoptLO} 
\end{equation}
 where $L_\mu \equiv \ln [(\mu+p_F)/M]$.
Like for the GN model one can see that the RGOPT extra terms bring in information from RG dynamics
through $g$ and $b_0$, to the otherwise trivial (free gas) perturbative result.
 Notice that in Eq. (\ref{rgoptLO}), assuming $\mu > m$ for most purpose below, the one-loop $\ln m$ terms of original 
 Eq. (\ref{PPTqcd}) have cancelled,
as a result of considering the vacuum contributions given by the first term in Eq. (\ref{PPTqcd}), such that 
the $\ln (m/M)$ is consistently replaced by a $\ln[(\mu+p_F)/M]$ with the same
coefficient\footnote{The very same cancellations occur in the 
original perturbative expression (\ref{PPTqcd}): this is not affected by RGOPT since the modification from 
(\ref{Lint}) modifies all $\ln m$ terms similarly.}.\\

Next, considering the  NLO RGOPT (i.e. taking $\delta\to 1$ in the $\delta^1$-order expansion of 
Eqs.~(\ref{Lint}), (\ref{sub})) and after some algebra the modified pressure (per flavor) reads:
\begin{eqnarray}
P^{RGOPT}_{1,f}(\mu)&=& P^{RGOPT}_{0,f}(\mu) 
- N_c \frac{m^4}{\left(4\pi\right)^2}  \left (\frac{\gamma_0}{b_0} \right ) \left ( \frac{1}{b_0\,g} \right)
+ m^4 \left (2\frac{\gamma_0}{b_0} -1\right ) s_1 \nonumber \\
&+& N_c \frac{m^2}{8\pi^2} \left( \frac{\gamma_0}{b_0}\right ) 
\left [ m^2 \left( 1 - 2L_\mu \right ) +2 \mu \, p_F  \right ] \nonumber \\ 
 &-& \frac{g d_A} {4\left(2\pi\right)^4} \left[ m^4 \left (\frac{1}{4} - 4 L_\mu +3 L_\mu^2 \right ) +
 \mu^2 \left(\mu^2+m^2 \right) +m^2\,\mu \, p_F \left(4 - 6 L_\mu \right) \right]   \;,
\label{rgoptNLO}
\end{eqnarray}
where $P^{RGOPT}_{0,f}(\mu)$ is given by Eq. (\ref {rgoptLO}).
Again, assuming $\mu>m$ for now on (except when explicitly mentioned below), we already
simplified the $\ln m$ terms at two-loop order, as those originating from the vacuum contributions cancel 
exactly with those similar terms of the medium parts, so that Eq.(\ref{rgoptNLO}) only depends
on the combination~\footnote{Those cancellations are however specific to
the one- and two-loop level: at higher orders $\ln m$ and  $\ln (\mu+p_F)$ appear independently,
due to more ``nested" divergences in the bare calculation \cite{paul3L}.} $L_\mu = \ln [(\mu+p_F)/M]$. 
The LO and NLO RGOPT pressure are now ready to be optimized to generate non-perturbative approximation results as
shown in the next section.

\section {Optimization procedure and numerical results }

\subsection{One-loop ($\delta^0$) RGOPT}
 Considering first the lowest ($\delta^0$) one-loop order result, let us recall that 
 the constraint from the reduced RG Eq. (\ref{red}) applied to the pressure, Eq. (\ref{rgoptLO}), has already been 
 used to fix the exponent of the interpolating Lagrangian, see 
 Eq. (\ref{aQCD}), such that by construction the pressure already satisfies $f_{RG}=0$ exactly (at this order).
 Consequently the arbitrary mass $m$ may be fixed only by using the MOP optimization equation:
 \begin{equation}
f_{MOP} = \frac{\partial {P}^{RGOPT}}{\partial m} \equiv 0 \;.
\label{mop}
\end{equation}
Considering first for simplicity solely the (one-loop) vacuum contribution in Eqs. (\ref{PPTqcd}) and (\ref{rgoptLO}) 
(reintroducing for this purpose consistently the $\ln m$ present at $\mu=0$), one obtains
\begin{equation}
 {\overline m}(\mu=0) =\Lambda_{\overline{\text{MS}}}\sqrt{e}\; ,
 \label{mLOT0}
\end{equation}
where $\Lambda_{\overline{\text{MS}}}=M e^{-\frac{1}{2gb_0}} $ is the one loop QCD
$\Lambda_{\overline{\text{MS}}}$ scale. Thus one obtains a non-perturbative mass, proportional to 
$\Lambda_{\overline{\text{MS}}}$, which is
exactly RG invariant.

Including next the one-loop in-medium contribution from Eq.~(\ref{rgoptLO}), we aim to use  
similarly the MOP Eq. (\ref{mop}) to now determine the $\mu$-dependent dressed mass ${\overline m}(\mu)$
(while the reduced RG equation is still automatically satisfied at this order for $\mu\ne 0$).
At finite densities Eq.~(\ref{mop}) is a little more involved due to the non-linear $m$-dependence 
from $p_F(m)=\sqrt{\mu^2-m^2}$ in Eq.~(\ref{rgoptLO}). Yet, after simple algebra the formal solution may be cast 
into a compact form:
\begin{equation}
 {\overline m}^2 = \mu^2 \left(\frac{\sqrt{1+4 c(\overline m,\mu,g)}-1}{2 c(\overline m,\mu,g)} \right) \;,
 \label{mop0}
\end{equation}
where\footnote{ Eq.(\ref{mop0}) suggests that $\overline m$ would be the solution of a simple quadratic equation 
if not for the nonlinear $m$ dependence entering $L_\mu= \ln[ (\mu+p_F)/M]$. In Eq.(\ref{mop}) we have selected the 
solution $m^2 >0$, while the other solution with $\sqrt{\cdots}\to -\sqrt{\cdots}$ is unphysical, giving always
$ m^2 <0$. }
\begin{equation}
c(m,\mu,g) = \left( \frac{1}{2 b_0 g} -\frac{1}{2} +L_\mu \right)^2\; .
\label{defc}
\end{equation}
  At this point we observe that the NLO subtraction $s_1$ in Eqs.(\ref{sub}),(\ref{s1}), 
while being strictly required for perturbative RG invariance only at two-loop
order, is formally a one-loop ${\cal O}(g^0)$ contribution.
It appears thus sensible to include this known information from next RG order, 
which is straightforward and provides not surprisingly a somewhat more realistic 
one-loop improved approximation. 
Accordingly for $s_1\ne 0$ the term $-1/2$ in Eq. (\ref{defc}) above is simply replaced by 
$-1/2-8\pi^2 s_1 =11/84$ (for $n_f=3$), which is the prescription used in the numerics below.\\

Eq. (\ref{mop0}) can be easily solved numerically but before doing that it is instructive to examine
some of its properties in more detail. One can see first that the coupling $g\equiv g(M)$ and the renormalization 
scale $M$ only appear in the combination $1/(2b_0 g) +L_\mu \simeq  1/(2b_0 g) +\ln(\mu/M)+\cdots$, 
where the dots designate $M$-independent terms. Therefore, recalling that the (exact) one-loop
running is defined as 
\begin{equation}
 g^{-1}(M) = g^{-1}(M_0) +2 b_0 \ln\left(\frac{M}{M_0}\right)\,,
 \label{g1L}
\end{equation}
for a reference scale $M_0$, 
it is immediate that Eq.(\ref{mop0}) does not at all depend on $M$. 
Likewise it is easy to see that the (LO) RGOPT pressure Eq.(\ref{rgoptLO}) is itself
exactly RG invariant at this one-loop order: formally replacing $m\to \overline m$ in its 
expression, $\overline m$ is RG invariant irrespectively of its numerical
value, and the explicit $g(M)$ and $M$ in Eq. (\ref{rgoptLO})
only appear in the very same $M$-independent combination  $1/(2b_0 g(M)) +L_\mu$. \\
For small coupling, the optimum mass $\overline m$ admits a perturbative expansion $\overline m^2\sim \mu^2 
(constant \times g +{\cal O}(g^2))$,
which has the expected form of an (in-medium) screening mass. Nevertheless, we insist at this point that $\overline m$ 
is not a physical mass, (and is not directly related to the Debye mass standard definition\cite{debye}), 
rather it represents an intermediate variational quantity whose sole purpose is to enter $P({\overline m},g,\mu)$, 
that {\em defines} the (optimized) physical pressure at a given order. In fact, except for very weak coupling, 
the first order expansion 
of $\overline m^2$ does not give a very good approximation of the exact $\overline m(\mu)$: indeed, 
instead of growing with no limits for arbitrary large coupling,
as the purely perturbative approximation would naively suggest, the exact solution in Eq. (\ref{mop0}) has the  
welcome property to be bounded, with $\overline m^2(g(M)) < \mu^2$ even for large $g(M)$ 
(therefore consistent with the basic assumptions of the in-medium contributions).
The numerical solution $\overline m$ at LO RGOPT from Eq. (\ref{mop0}) as a function of $\mu$ is illustrated
in Fig. 2, which among other features evidently confirms its exact
scale invariance properties.

\subsection{NLO two-loop ($\delta^1$) RGOPT: in-medium contribution}\label{rgonlo}
At NLO ${\cal O}(g)$, it turns out that Eqs.(\ref{mop}) and (\ref{red}) 
do not have real solutions for arbitrary chemical potential values. As already discussed above in Sec.\ref{secGN} 
this is expected to happen in general, starting at NLO order, due to non-linear dependences in the mass, if 
we insist to solve those equations exactly. 
Therefore, one could try less rigidly to solve the sole complete RG equation, Eq. (\ref{RG}), for $\overline m(g)$, 
taking then for $g$ more conservatively the purely perturbative two-loop running coupling. 
Unfortunately only non real solutions appear also in this case, if the equation is solved for exact $m(g)$. 
Nevertheless this situation can be remediated, at the price of introducing one extra parameter,
to be fixed however by a well-defined prescription.
Following Ref. \cite {JLalphas}, the idea is to modify the perturbative coefficients, expecting
in this way to recover real solutions. But the modification should not be arbitrary, and should be RG compatible, 
so a presumably sensible prescription is to perform a (perturbative) renormalization scheme change (RSC). 
With a little  insight, since one is mainly concerned with mass optimization, a simplest RSC can 
be defined by modifying only the mass parameter according to
\begin{equation}
m \to m^\prime ( 1+ B_1 g + B_2 g^2 +...) \,,
\label{RSC}
\end{equation}
where the $B_i$ parametrize arbitrary scheme changes from the original ${\overline{\text{MS}}}$-scheme 
\footnote{Eq.(\ref{RSC}) has also the welcome property that it does not affect the definition of 
the reference QCD scale $\Lambda_{\overline{\text{MS}}}$, in contrast with a similar perturbative 
change on the coupling, see \cite{JLalphas} for details.}.  
For an exactly known function of $m$ and $g$ Eq. (\ref{RSC})  would
just be a change of variable not affecting physical
results. While for a perturbative series truncated at
order $g^k$, different schemes differ formally by
remnant term of order ${\cal O}(g^{k+1})$, such that the difference between two schemes 
is expected to decrease at higher orders for sufficiently weak coupling value. Now 
since we aim to solve optimization equations for ``exact" $m$ and $g$ dependence, Eq. (\ref{RSC}) 
actually modifies those purposefully, when now considered as constraints for the arbitrary mass $m^\prime$. 
Furthermore we vary only one RSC parameter consistently at the same perturbative order, 
such that the relevant form of Eq. (\ref{RSC}) is $m\to m^\prime (1+ B_2 g^2)$: thus 
upon re-expanding to order-$g$ one can easily see that the net RSC modification to the pressure 
is to add the extra term  $- 4g m^4 s_0 B_2 $ at two-loop order-$g$ (and simply renaming $m^\prime \to m$ afterwards 
the mass parameter to be determined to avoid 
excessive notation changes).

Clearly a definite prescription is required in order to fix $B_2$. 
Accordingly, one requires \cite{JLalphas} the RSC to give the real $\overline m$ solution the {\em closest } 
to the original $\overline{\text{MS}}$-scheme: that is mathematically expressed by 
requiring the ``contact" of the two curves (i.e. collinearity of the vectors tangent) 
parametrizing the relevant MOP and RG equations, considered as functions of $(m, g)$:
\begin{equation}
f_{RSC}= \frac{\partial f_{RG}}{\partial g} \frac{\partial f_{MOP}}{\partial m} - 
\frac{\partial f_{RG}}{\partial m} \frac{\partial f_{MOP}}{\partial g} \equiv 0\,,
\label{contact}
\end{equation}
where $f_{MOP}$ and $f_{RG}$ are given respectively by Eqs. (\ref{mop}) and (\ref{red}) (applied here to the QCD pressure). 
Thus, Eq. (\ref{contact}) provides an extra constraint which completely
fixes the additional RSC parameter $B_2$. Moreover, one expects the RSC to remain reasonably perturbative, i.e. $B_2$
to be moderate, which may be verified a posteriori by inspecting that $B_2 \, g^2 \ll 1$.   \\

 As a first simple illustration, let us consider only the vacuum contribution 
at $\mu=0$. Applying Eq. (\ref{contact}), in conjunction with the MOP Eq.(\ref{mop}) 
and taking for $g$ the two-loop perturbative Eq.(\ref{g2L}), and for a typical value $M\simeq 1\, {\rm GeV}$, 
one then obtains ${\overline {B}}_2 \simeq -0.00224$,  
$\ln ({\overline m}/M)\simeq -0.331$, giving $\overline m \simeq 700$ MeV 
$\simeq 2.1\,\Lambda_{\overline{\text{MS}}}$,
which may be compared to the LO RGOPT result Eq.(\ref{mLOT0}).  Note that $\alpha_S(M=1\,{\rm GeV})\simeq 0.42 $,
but $|\overline B_2 g^2|\sim 0.06$ is a very moderate deviation from original
$\overline{\text{MS}}$-scheme.

We can now numerically optimize the NLO RGOPT pressure Eq.~(\ref{rgoptNLO}) 
including the in-medium contribution with $\mu \ne 0$,  
adopting the RSC additional prescription to recover real $\overline m(\mu)$ solutions 
for arbitrary $\mu$ values. 
Actually, rather than solving the full RG, as a numerically simpler variant we solve 
the MOP equation (\ref{mop0}) at two-loop order for $\overline m(g)$, taking for $g$ the purely
perturbative running coupling, together with the RSC equation (\ref{contact}) to fix 
${\overline B}_2(\mu)$ at NLO. (Clearly the optimal RSC parameter $\overline B_2$ will now be a nontrivial 
function of the chemical potential $\mu$, consistently determined by the optimization procedure). 
At two-loop the analytical expression of the MOP, Eq.(\ref{mop}), is more involved than its one-loop
analogue, so that the algebra leading to the solution Eq.(\ref{mop0})
is not as easy but it can be readily solved numerically.\\ 
In order to compare with the results given in 
Refs. \cite {paulPRL,paul3L} we will consider the scale variation $\mu \le M \le 4\mu$ besides the ``central" scale 
$M=2\mu$.
The exact two-loop (2L) running coupling, analogue of the one-loop Eq.(\ref{g1L}), is obtained by solving for $g(M)$
the relation
\begin{equation}
%
\ln \frac{M}{ \Lambda_{\overline{\text{MS}}} } = \frac{1}{2b_0\, g} +
\frac{b_1}{2b_0^2} \ln \left ( \frac{b_0 g} {1+\frac{b_1}{b_0} g} \right) ,
\label{g2L}
\end{equation}
for a given $\Lambda_{\overline{\text{MS}}}$ value (this also defines the (two-loop order) 
$\Lambda_{\overline{\text{MS}}}$ in our normalization conventions).
Equivalently to giving a $\Lambda_{\overline{\text{MS}}}$ value one can give a $g(M_0)$ at some reference scale,
$M_0$.  Here, we have chosen $\alpha_s(M_0= 1.5 \, {\rm GeV})=0.326$ to compare precisely with the values adopted in 
Ref. \cite{paulPRL}: with (\ref{g2L}) this corresponds to $\Lambda_{\overline{\text{MS}}}\simeq 0.335\,{\rm GeV} $, 
a value indeed very close to the present world average \cite{PDG2018}.

\begin{figure}[htb]
 \centerline{ \epsfig{file=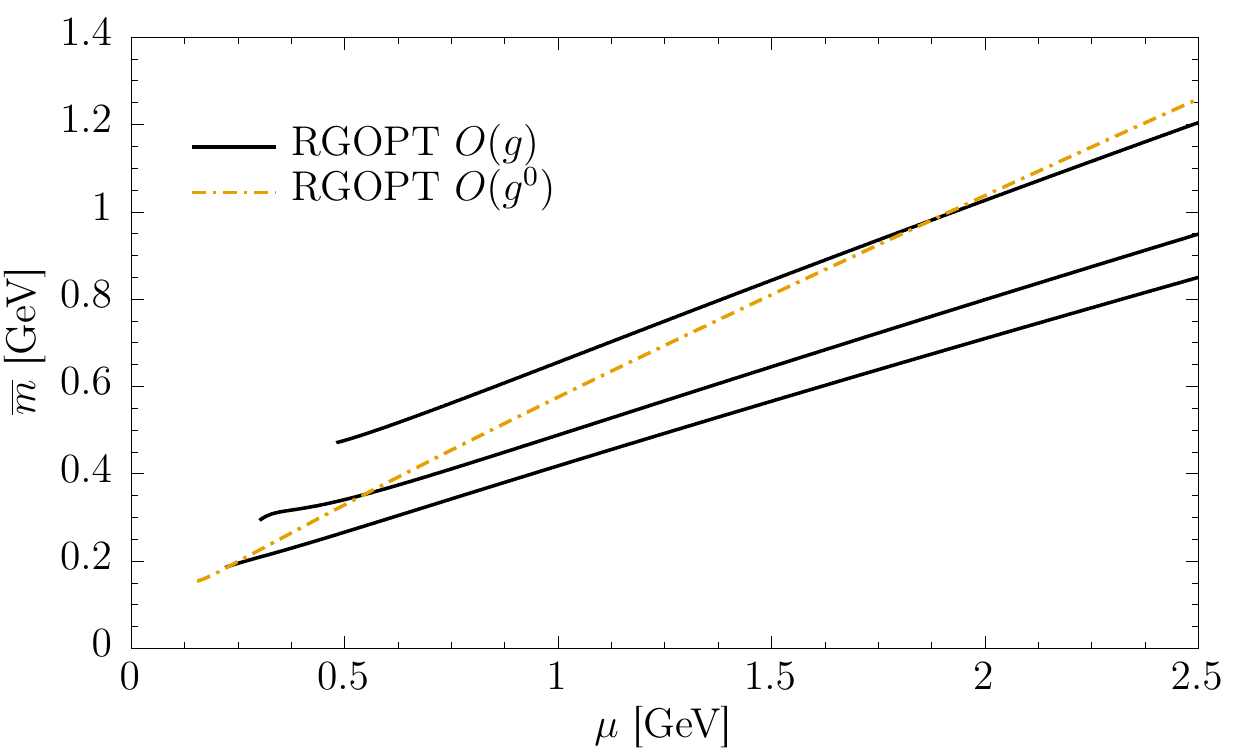,width=0.4\linewidth,angle=0}}
 \caption{ The optimized mass as a function of the chemical potential at $M=\mu, 2\mu$ and $4\mu$ at 
 order-$g^0$ (dot-dashed) and order-$g$ (continuous). For the latter the upper curve corresponds 
 to  $M=\mu$, the central curve to $M=2\mu$, and the lower curve to $M=4\mu$.  }
 \label{Fig2}
\end{figure}

\begin{figure}[htb]
 \centerline{ \epsfig{file=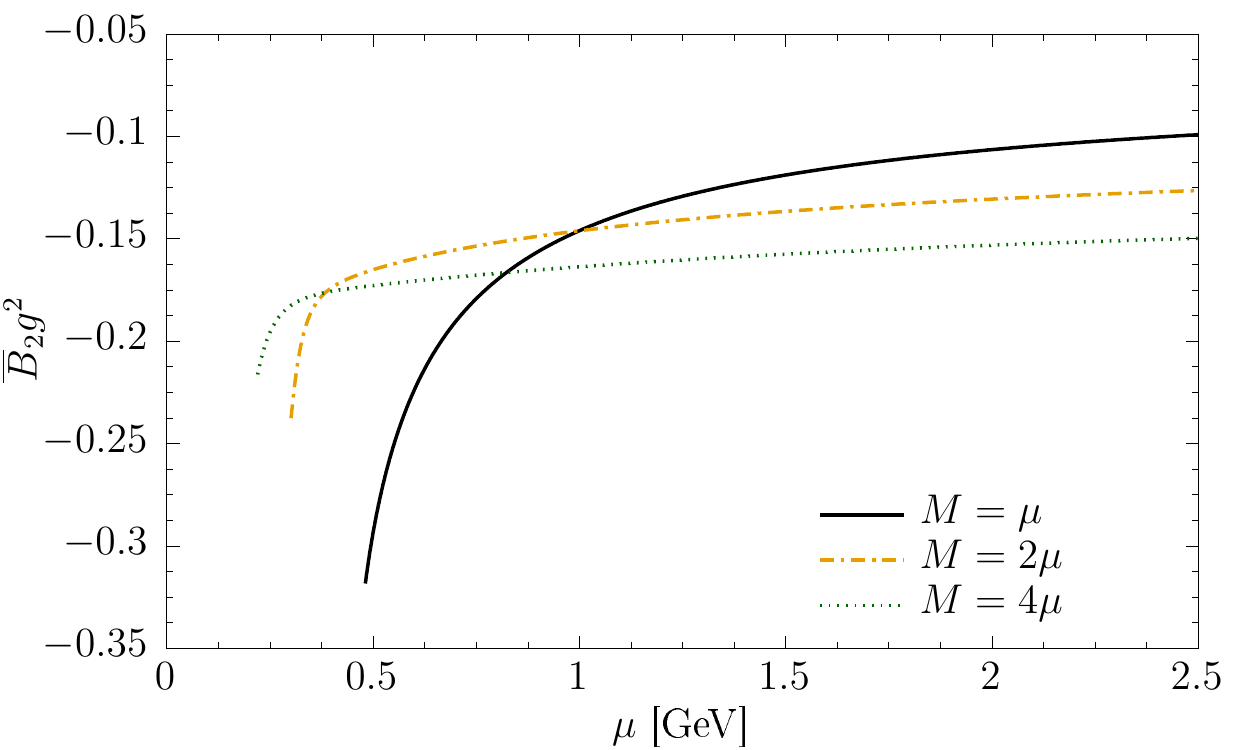,width=0.4\linewidth,angle=0}}
 \caption{ The optimized RSC $\overline B_2(\mu) g^2(\mu)$ quantity as a function 
 of the chemical potential at $M=\mu, 2\mu$ and $4\mu$ at order-$g$.   }
 \label{Fig3}
\end{figure}

The results for the optimized NLO mass $\overline m$ as functions of $\mu$ and for different renormalization scale
choices are shown in Fig. \ref {Fig2} where they are also compared with the LO one-loop $\overline m$ value.
One can see that, as already explained above, 
${\overline m}(\mu)$ is exactly RG invariant at LO, because it 
only involves the scale invariant combination $[2b_0 g(M)]^{-1}+\ln (\mu/M)$. 
In contrast the NLO ${\overline m}(\mu)$ definitely displays a residual scale dependence: 
even for the exact two-loop running, Eq.(\ref{g2L}), the latter is not very surprisingly 
no longer exactly ``matched" by the optimized NLO RGOPT mass. 
We will illustrate below that  the RGOPT pressure, which represents the actual physical observable,
shows a more moderate residual scale dependence. More generally
the RGOPT construction only guarantees that the optimization does not spoil the {\em perturbative} 
RG invariance of the physical quantity considered, 
that means up to remnant scale-dependent terms of higher order ${\cal O}(g^{k+1})$, if 
the original perturbative expression is available at order $g^k$. \\
Fig. \ref{Fig3} illustrates the corresponding values of the RSC parameter combination $B_2(\mu) g^2(\mu)$, 
thus quantifying the departure from  $\overline{\text{MS}}$-scheme. One can see that RSC remains reasonably
perturbative, although the value of $|B_2(\mu)|$ needed to recover real solutions are increasing 
rapidly for smaller $\mu$ values for the lower renormalization scale $M=\mu$ (not surprisingly
since in this region the running coupling $g(M)$ becomes dangerously large).\\ 

One is now in position to compute thermodynamical observables, such as the pressure and the quark number 
density which here will be respectively normalized by the equivalent  massless free gas quantities 
$P_{fg}$ and  $\rho_{fg}$. These quantities, per flavor,  are respectively
\begin{equation}
P_{fg}= N_c\frac{\mu^4}{12\pi^2} \,,
\end{equation}
and
\begin{equation}
\rho_{fg}= N_c\frac{\mu^3}{3\pi^2} \,.
\end{equation}
\begin{figure}[htb]
 \centerline{ \epsfig{file=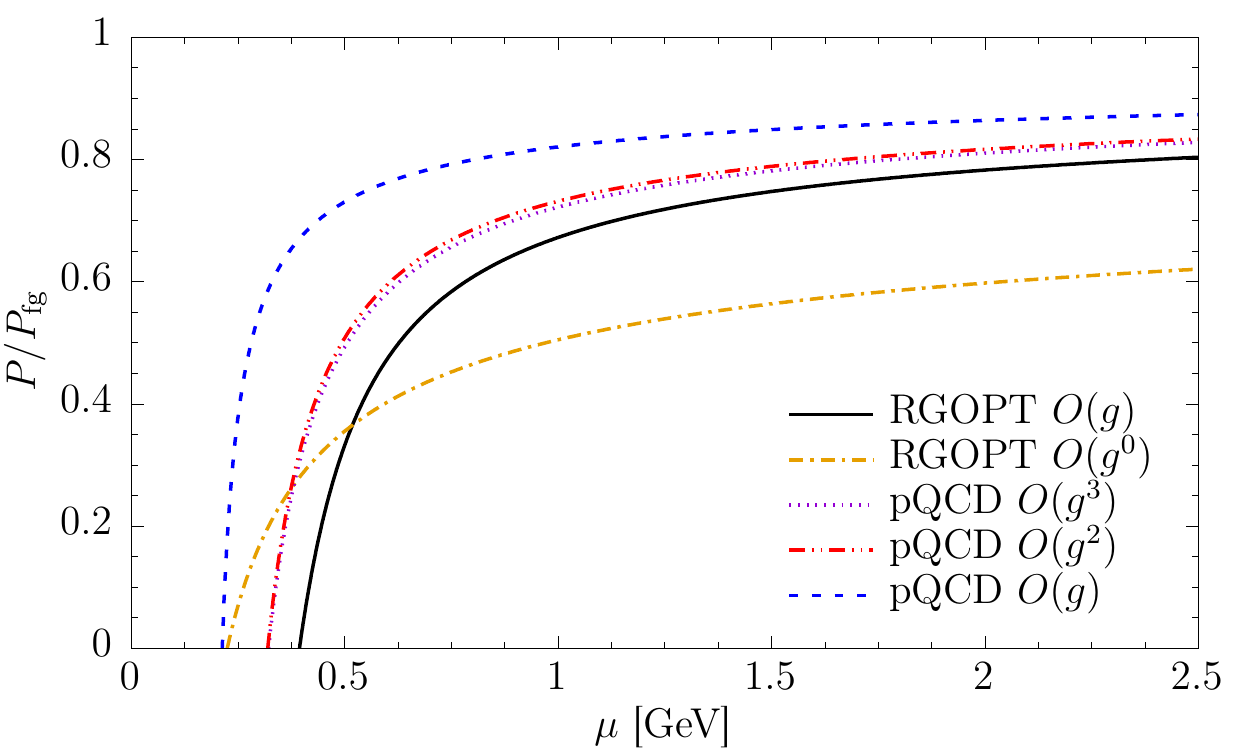,width=0.4\linewidth,angle=0}}
 \caption{ The normalized pressure as a function of the chemical potential at the central scale 
 $M=2\mu$. pQCD results at orders $g$, $g^2$, and $g^3$ (LL term) are compared with 
 the RGOPT results at orders $g^0$ (one loop) and $g$ (two loop). }
 \label{Fig4}
\end{figure}

\begin{figure}[htb]
 \centerline{ \epsfig{file=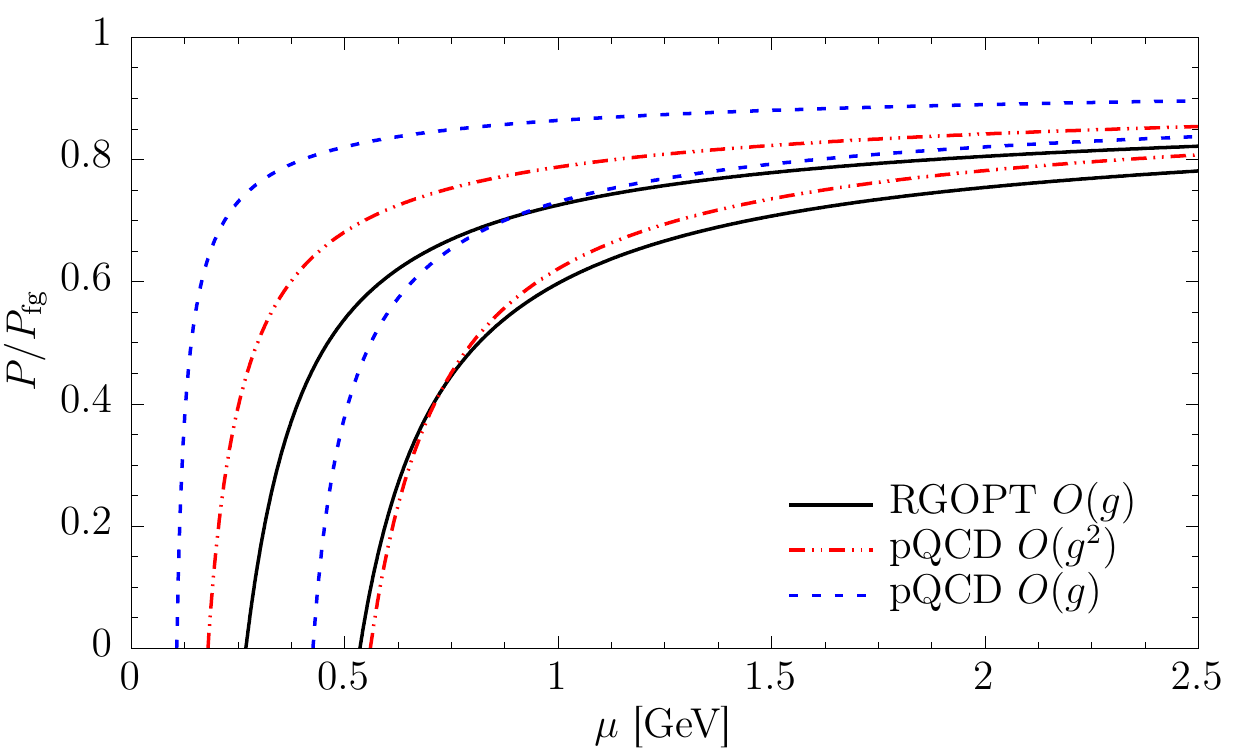,width=0.4\linewidth,angle=0}}
 \caption{ The normalized pressure as a function of the chemical potential. 
 pQCD results at NLO order-$g$ and NNLO order-$g^2$ are compared with RGOPT at
 NLO order-$g$. In each case the upper curve corresponds to $M=4\mu$ 
and the lower curve to $M=\mu$.  }
 \label{Fig5}
\end{figure}
\begin{figure}[h!]
 \centerline{ \epsfig{file=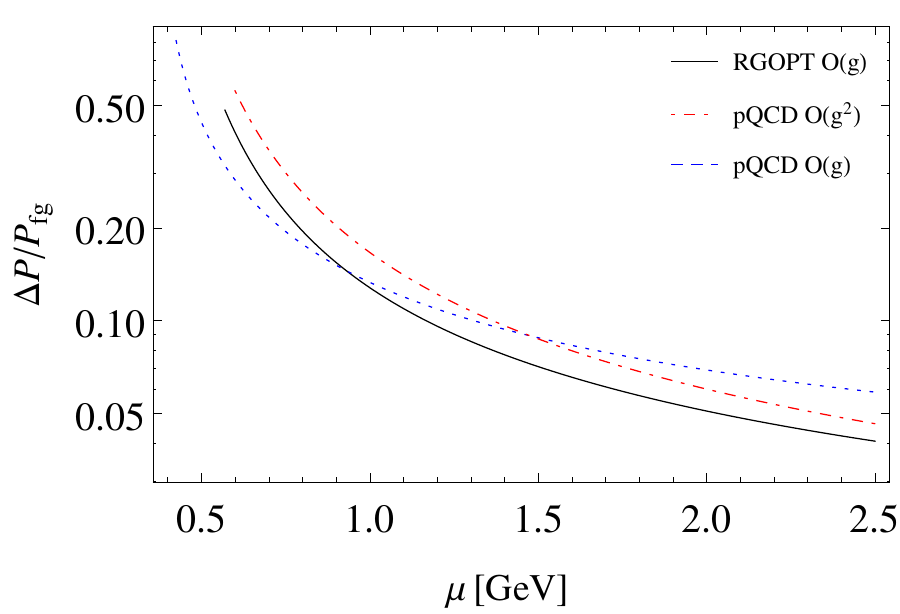,width=0.4\linewidth,angle=0}}
 \caption{ The remnant scale dependences defined by the differences 
 $\Delta P/P_{fg} \equiv (P(M=4\mu)-P(M=\mu))/P_{fg}$ 
 of (normalized) pressures, as functions of the chemical potential $\mu$. 
 pQCD results at NLO order-$g$ and NNLO order-$g^2$ are compared with RGOPT at NLO order-$g$.}
 \label{Fignew}
\end{figure}

Let us then compare the ${\cal O}(g^0)$ and ${\cal O}(g)$ RGOPT results with the
pQCD predictions  at ${\cal O}(g)$, ${\cal O}(g^2)$\cite{paul3L}, as well as the most recent 
${\cal O}(g^3 \ln^2 g)$~\cite{paulPRL}. For completeness, we recall that the relevant pQCD 
expression is \cite{paulPRL} 
\begin{equation}
\frac{P^{pQCD}}{P_{fg}} = 1 - \frac{2}{\pi} \alpha_S(M) -\alpha^2_S(M) \left \{  0.303964 \ln \alpha_S(M) 
 +\left [ 0.874355 + 0.911891 \ln \left (\frac{M}{\mu} \right ) \right] \right \} \nonumber \\
 - 0.266075 \alpha^3_S(M) \ln^2\alpha_S \; .
 \label{pQCD}
\end{equation}
In Fig. \ref {Fig4} we show the normalized pressure 
predicted by the different order approximations at the central scale choice $M=2\mu$ as adopted in Ref.~\cite{paulPRL}. 
The first thing to remark is that the RGOPT 
produces a non-trivial result already at order-$\delta^0 g^0$, but converging quite slowly to the free gas 
result as the quark chemical potential increases. Nevertheless this can already be seen as an improvement 
since at this same order $g^0$ the pQCD result for the normalized pressure would trivially be equal to 
the unity, i.e., the free gas limit. In fact the lowest order RGOPT cannot be expected to be a very realistic
approximation in general, because it only relies on lowest order RG quantities, while the  pressure dependence  
is essentially like the free gas one.
The efficient resummation properties of RGOPT become more evident  
when one compares its result at NLO, order-$g$, with the pQCD ones at the same NLO order, since the figure 
shows that the NLO RGOPT pressure actually appears in much better agreement with the 
{\em higher order} perturbative $g^2$ and $g^3 \ln^2 g$ predictions. 
Next we also analyze how the different approximations perform when the arbitrary renormalization scale is varied 
in the range $\mu \le M \le 4\mu$, as in Ref. \cite{paul3L} where the scale dependence of pQCD results at orders
$g$ and $g^2$ have been analyzed~\footnote{In the original study~\cite{paul3L}
a quite common approximate form of (\ref{g2L}) was rather used, truncating terms beyond ${\cal O}(\ln L/L^2)$,
with $L\equiv \ln(M^2/\Lambda^2_{\overline{\text{MS}}})$. Here 
we compare the scale dependence by adopting the same exact two-loop running coupling (\ref{g2L}) for all 
approximations, that tends to very slightly decrease the remnant scale uncertainty for all cases.}. 
The results are compared in Fig. \ref {Fig5},  where the RGOPT appears to moderately improve 
the scale uncertainty, at least in the range $\mu \gtrsim 1$ GeV, as compared with 
the same perturbative order $g$. \\
To assess more precisely the remnant scale dependence we plot
in Fig. \ref{Fignew} the difference of the (normalized) pressures  
$\Delta P/P_{fg} \equiv (P(M=4\mu)-P(M=\mu))/P_{fg}$  
as function of $\mu$, for the three approximations illustrated in Fig. \ref{Fig5}. 
The NLO RGOPT remnant scale dependence
is moderately but clearly improved as compared to NLO pQCD for $\mu \gtrsim 0.9$ GeV 
(giving $\sim 25\%$ improvement e.g. for $\mu\simeq 2$ GeV), while the
NLO pQCD scale dependence appears somewhat smaller in the lower $\mu$ range $0.5 \lesssim \mu \lesssim 0.9$ GeV. 
Notice also that the NNLO pQCD pressure has a smaller scale dependence than NLO pQCD 
in a narrower and more perturbative range $\mu \gtrsim 1.5$ GeV. In contrast
the RGOPT scale uncertainty is clearly better than the NNLO pQCD one in the full relevant $\mu$ range.
We remark however that the smaller remnant
dependence of NLO pQCD within the low-$\mu$ window ($0.5 \lesssim \mu \lesssim 0.9$ GeV)
is merely a side effect of the NLO pQCD pressure 
dropping towards zero at lower $\mu$ values than the two other approximations, as is clear from Fig. \ref{Fig5}.
Indeed not surprisingly all three approximations exhibit a rapidly growing scale dependence
for $\mu$ values approaching the region where $P(M\simeq \mu)$ rapidly drops 
towards zero~\footnote{In Fig. \ref{Fignew} the three curves consistently start 
at their respective minimal $\mu_{min}$ values, defined 
such that $P(M\ge \mu_{min})\ge 0$, compare with Fig. \ref{Fig5}.}.  
But Fig. \ref{Fignew} also shows that the {\em maximal} remnant dependence reached at the respective 
$\mu_{min}$ values is smaller for the RGOPT than for NLO and NNLO pQCD. 
In any case one should keep in mind that, due to the adopted common renormalization scale choice 
$g(M ={\cal O}(\mu))$, none of the approximations is much reliable 
in the nonperturbative region where $P/P_{fg}(M\simeq \mu) \ll 1$ due to large coupling  
(note, e.g., that $\mu < 0.8$ already corresponds to $\alpha_S(M=\mu) > 0.5 $ using Eq.(\ref{g2L})). 
We thus conclude that, within the $\mu$ range where all the approximations are very
reliable perturbatively, the NLO RGOPT remnant scale uncertainty is 
moderately but clearly improved in relative comparison to both NLO and NNLO pQCD (considering also that 
standard pQCD at $T=0, \mu \ne 0$ has anyway less severe scale dependence issues than in the high $T$ regime).\\

In principle, we could also include in our NLO RGOPT analysis the NNLO $g\, m^4\, s_2$ subtraction term
of Eq.(\ref{sub}), since being formally of order-$g$, similarly to what was done at LO RGOPT 
(see the discussion after Eq.(\ref{defc})). The $s_2$ expression is available from \cite{JLcond} 
and clearly incorporates additional
RG dependence from next (three-loop) RG order. However we have checked that considering $s_2\ne 0$ 
at NLO scarcely changes our results (in contrast with the LO pressure where $s_1 \ne 0$ has
a sizeable impact). In particular the scale dependence is not visibly affected,
which signals that a reasonable stability has been reached at NLO order.

\begin{figure}[h!]
 \centerline{ \epsfig{file=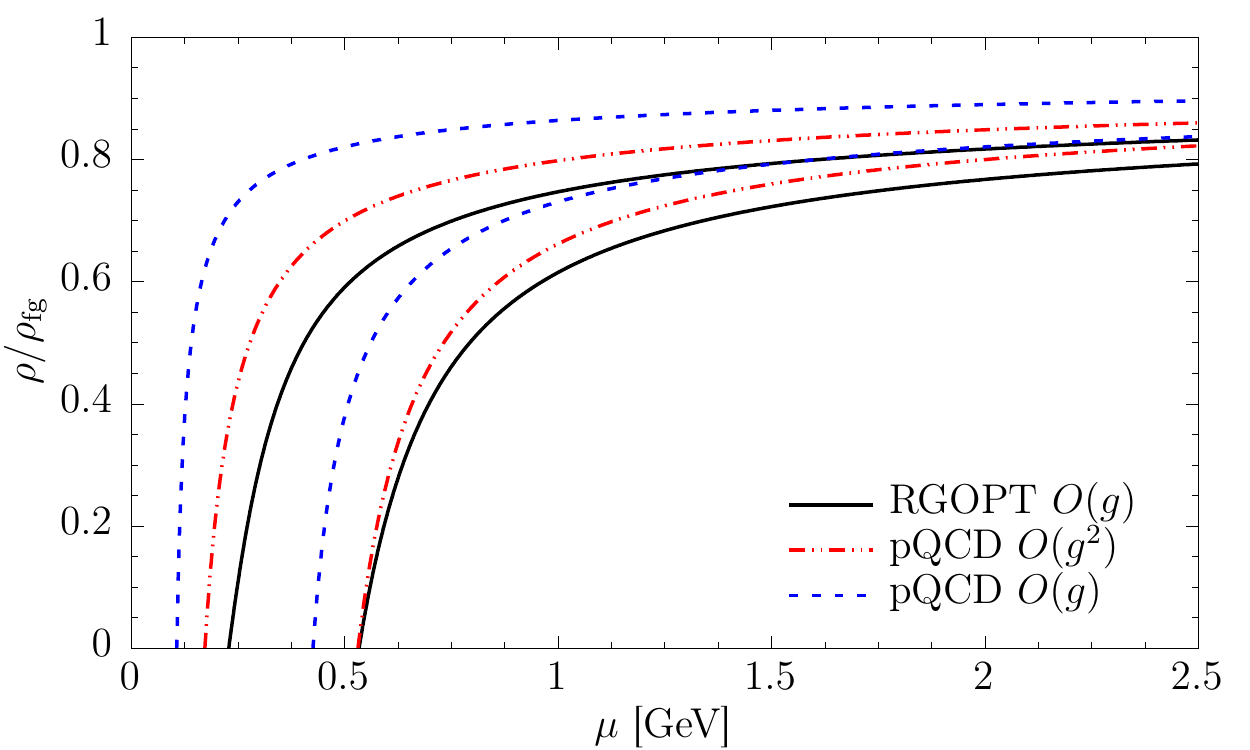,width=0.4\linewidth,angle=0}}
 \caption{ The quark number density as a function of the chemical potential. 
 pQCD results at NLO order-$g$ and NNLO order-$g^2$ are compared with the RGOPT at 
 NLO order-$g$. In each case the upper curve corresponds to $M=4\mu$ 
 and the lower curve to $M=\mu$.  }
 \label{Fig6}
\end{figure}

In Ref. \cite {paul3L} the authors also analyzed the predictions for the quark 
number density:
\begin{equation}
 \rho(\mu) \equiv \frac{d P (\mu)}{d\,\mu} \; ,
\label{rho}
\end{equation}
 up to NNLO pQCD.
Their results are reproduced and compared with our RGOPT predictions 
in Fig. \ref {Fig6}. As in the case of the pressure a noticeable  (but moderate) 
decrease of the scale ``uncertainty" 
band occurs for the NLO RGOPT density (while the LO RGOPT results 
are again exactly RG invariant for the same reasons than the LO pressure). 
However the RGOPT scale dependence improvement is less pronounced
than for the pressure, which can be traced to our use of the standard running coupling 
(having renounced, as explained above, to the more complete optimization of $g$ and $ m$, 
due to non real and involved solutions). Indeed, for dense matter the imperfectly balanced scale dependence, 
from the contribution of the running $g(M)$, tends to be enhanced as compared to the pressure 
since taking $g(M\sim \mu)$ to obtain $\rho(\mu)$ in Eq.(\ref{rho}) involves a contribution $\propto \partial_M g(M)$ 
on top of the explicit derivative  $\partial_\mu P$ term. 

\subsection{A simpler alternative NLO RGOPT prescription}
\begin{figure}[h!]
 \centerline{ \epsfig{file=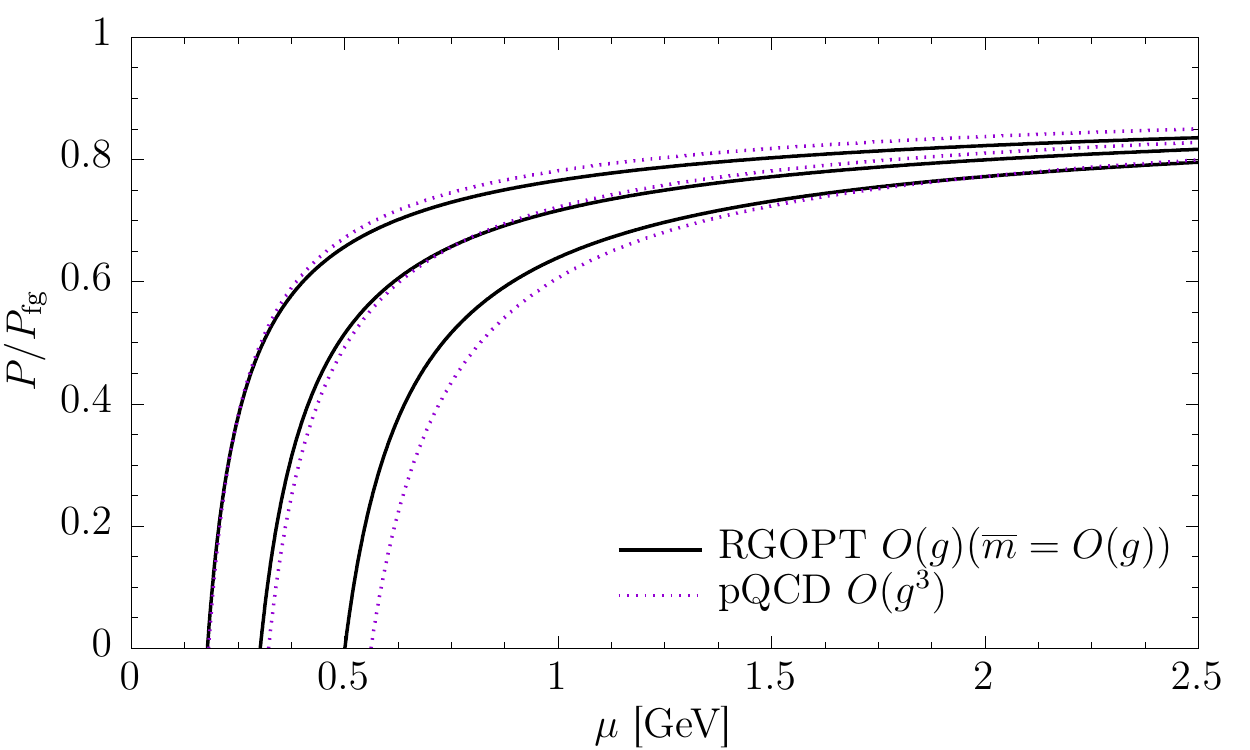,width=0.4\linewidth,angle=0}}
 \caption{The normalized pressure as a function of the chemical potential. 
 pQCD results from Eq.(\ref{pQCD}) at NNNLO including the contribution $g^3 \ln^2 g$ 
 (blue dotted curves) compared with 
 the simpler alternative NLO RGOPT (black continuous curve). 
 In each case the upper curve corresponds to $M=4\mu$, the central curve to 
 $M=2\mu$, and the lower curve to $M=\mu$.  }
 \label{Fig7}
\end{figure}

While the results in Figs. \ref{Fig4} and  \ref{Fig5} clearly show a better agreement of 
NLO RGOPT 
with the state-of-the-art perturbative results, it may be regarded rather unsatisfactory to have to deal 
with the somewhat more involved RGOPT NLO prescription, that implies 
the additional constraint from RSC Eq.(\ref{contact}) to be altogether numerically solved to restore real 
solutions. 
Could we find a simpler and more transparent prescription, while still capturing the main features of the RGOPT approach? 
Indeed, a much simpler alternative that surely recovers a real $\overline m$ solution  is 
simply to renounce to solving the RG or MOP equations {\em exactly}, by approximating 
the latter in a more perturbative fashion. (This, however, certainly looses a part of the
resummation properties embedded in the ``exact" solution, such that a slight degradation of 
the remnant scale dependence is to be expected). To explore this alternative we consider 
the full RG equation (\ref{RG}),  in order to incorporate the most complete and consistent NLO RG content, but we
approximate crudely its solution to its first perturbative (re)expansion order. Similarly as in the LO
case, noting first that Eq.(\ref{RG}) would give
a simply quadratic equation for $ m^2$ in absence of the extra nonlinear $m$-dependence from $p_F$, 
this perturbative solution is simple:\footnote{In Eq.(\ref{msimp}) the factors $9/7$, $21/64$, $\sqrt{43}$
are simply $n_f=3$ values of the specific combinations of RG coefficients $b_i$, $\gamma_i$ appearing in this 
expression.  The explicitly scale-dependent term $\ln \mu/M$ only appears at next $g^2$-order. 
Note also that we eliminated the other solution, with $+\sqrt{\cdots}$, as it violates the necessary 
consistency $\overline m^2 \le \mu^2$ even for moderate $g$, and also does not fullfill the 
perturbative matching, $\ln \mu/\overline m \sim 1/(2b_0 g)$, for $\mu \gg \overline m$.}
\be
\overline m^2 =\frac{9}{7\pi^2} \mu\,p_F \left (1-\sqrt{1-\frac{21}{64}\frac{\mu^2}{p^2_F}} \;\right )\,g  
+{\cal O}(g^2)
= \frac{9}{7\pi^2} \left( 1 -\frac{\sqrt{43} }{8} \right )\,g\,\mu^2 +{\cal O}(g^2) \,.
\label{msimp}
\ee
Now, inserting this $\overline m$ expression into the NLO RGOPT pressure expression Eq. (\ref{rgoptNLO}), with
the running $g\to g(M)$ as previously from Eq. (\ref{g2L}), gives the results shown in Fig. \ref{Fig7}, 
which are compared
with pQCD at NNLO including the four-loop (LL) results from Eq. (\ref{pQCD}) 
(originally obtained in Ref. \cite{paulPRL}). 
We illustrate also the scale dependence for the range $\mu \le M\le 4\mu$ for the two expressions. 
One sees the quite remarkable agreement
for the central scale choice $M=2\mu$ (more precisely with less than $\sim 1.5\%$ differences for 
any $\mu>0.6$ GeV), while 
the RGOPT scale ``uncertainty" range is still slightly better even for this rather crude approximation. Concerning 
the scale dependence band of pQCD including the highest available perturbative order result, Eq. (\ref{pQCD}), it
hardly displays a visible difference with the sole NNLO, order-$g^2$, perturbative pressure as studied in 
Ref. \cite{paul3L}. But the net effect of the highest order last term of Eq. (\ref{pQCD}), being negative, is 
 to shift down (very slightly) the values of the pressure for given $\mu$ and $M$ values.
Examining Fig. \ref{Fig4} we further observe that going from NLO to NNLO pQCD there is a more pronounced decrease
of the pressure for given $\mu$ values (which is clear from the globally negative 
NNLO ${\cal O}(g^2)$ terms in Eq.~(\ref{pQCD})). Now, in Fig. \ref{Fig4} the {\em exact} NLO RGOPT
pressure values are sensibly lower than the other approximations, while in contrast the 
approximate NLO RGOPT pressure, obtained with the perturbative $\overline m$ Eq.(\ref{msimp}), agrees
quite neatly with Eq.(\ref{pQCD}). Accordingly one may hint from those comparisons that  
the ``exact" NLO RGOPT result may be a more precise approximation than Eq.~(\ref{pQCD}) 
to the even higher order perturbative pressure values. 

\section{Conclusions}

In this work we have performed the first application of the RGOPT resummation to QCD  when a control parameter, 
such as the chemical potential, is present. As discussed this technique generates non-perturbative 
approximations with consistent RG properties in a region of the QCD phase diagram which is currently unavailable to LQCD simulations. 
Our results have been  compared to the state-of-the-art pQCD predictions that include a $\alpha_s^3 \ln^2 \alpha_s$
contribution. We have confirmed in this in-medium application the generic property that at lowest
one-loop order this technique already captures non-trivial and RG invariant
results for the pressure and the quark number density.
Although numerically these lowest order results 
are a poor approximation in general, and 
converge quite slowly to the free gas result as $\mu$ increases, they exhibit the more efficient 
RGOPT resummation since, at this same order,  the pQCD prediction is trivial. At NLO order-$g$ (two loop level) 
and $M=2\mu$ the RGOPT results appear to be a very good approximation as they show a much better agreement 
with the perturbative higher orders ${\cal O}(\alpha_s^3 \ln^2 \alpha_s)$ 
than pQCD at the same order. Scale variations in the range $M=\mu-4\mu$ also show that the method reduces the 
scale uncertainties (although moderately at two-loop order) as compared to pQCD, 
which is important as far as EoS suitable to describe 
neutron stars  are concerned. \\
The scale uncertainty improvement from RGOPT thus 
appears less spectacular than for other models explored at two-loop orders at $T\ne 0, \mu=0$ 
compared with standard perturbation and HTLpt \cite{prlphi4,prdphi4,nlsm}. But this is merely due to 
the fact that standard pQCD at $T=0, \mu \ne 0$ has less severe remnant scale dependence issues (as already
noted in Ref.\cite{paul3L}) than most other models have in the high $T$ regime.
In contrast the NLO RGOPT scale uncertainty appears more similarly moderate in both regimes.
As discussed in the text and in other applications (see, eg, Ref. \cite {prdphi4}) 
the appearance of a residual (mild in most cases) scale dependence is unavoidable within the RGOPT beyond LO. 
But it is also clear \cite{prdphi4} that since RGOPT maintains by construction the most possible 
of (perturbative) RG invariance, generically the scale uncertainty bands observed at NLO should further shrink by 
considering the NNLO, ${\cal O}(g^2)$, which should also provide a priori more accurate numerical results. We remark 
that by combining the three loop vacuum contributions of Ref.  \cite {JLcond} with the in-medium contributions of 
Ref. \cite {paul3L} this is a feasible, although technically more involved analysis (regarding optimization), 
that we intend to address in a future investigation. Regarding the present  application, where 
only massless quarks have been considered, our results indicate that this RG-consistent 
resummation method is suitable to treat dense and cold QCD. Note also that it can be easily extended
to determine more realistic EoS (e.g., including massive quarks) which aim to describe neutron stars. Finally, the 
RGOPT interpolation should be extended to the gluonic sector for a more complete description specially
when considering high temperature effects \cite{gluons}.

\acknowledgments
 
This  work  was  financed  in  part  by    Coordena\c c\~{a}o  de
Aperfei\c coamento  de  Pessoal  de  N\'{\i}vel  Superior  - (CAPES-Brazil)  -
Finance  Code  001 and   by INCT-FNA (Process No.  464898/2014-5).  T.E.R.  
acknowledges Conselho Nacional de Desenvolvimento Cient\'{\i}fico e Tecnol\'{o}gico (CNPq-Brazil) and  Coordena\c c\~{a}o  de
Aperfei\c coamento  de  Pessoal  de  N\'{\i}vel  Superior   (CAPES-Brazil) for   PhD grants at different periods of time.  
M.B.P. and J.-L.K. thank Paul Romatschke for discussions and the Department of Physics at UC Boulder, 
where this work was completed, for the hospitality.

\bibliographystyle{apsrev4-1} 
\bibliography{bibliography}

\end{document}